*Systematic Review*

# Innovative Speech-Based Deep Learning Approaches for Parkinson's Disease Classification: A Systematic Review

Lisanne van Gelderen and Cristian Tejedor-García *

Centre for Language Studies, Radboud University, 6500 HD Nijmegen, The Netherlands; lisanne.vangelderen@ru.nl
* Correspondence: cristian.tejedorgarcia@ru.nl

**Abstract:** Parkinson's disease (PD), the second most prevalent neurodegenerative disorder worldwide, frequently presents with early-stage speech impairments. Recent advancements in Artificial Intelligence (AI), particularly deep learning (DL), have significantly enhanced PD diagnosis through the analysis of speech data. Nevertheless, the progress of research is restricted by the limited availability of publicly accessible speech-based PD datasets, primarily due to privacy concerns. The goal of this systematic review is to explore the current landscape of speech-based DL approaches for PD classification, based on 33 scientific works published between January 2020 and March 2024. We discuss their available resources, capabilities, and potential limitations, and issues related to bias, explainability, and privacy. Furthermore, this review provides an overview of publicly accessible speech-based datasets and open-source material for PD. The DL approaches identified are categorized into end-to-end (E2E) learning, transfer learning (TL), and deep acoustic feature extraction (DAFE). Among E2E approaches, Convolutional Neural Networks (CNNs) are prevalent, though Transformers are increasingly popular. E2E approaches face challenges such as limited data and computational resources, especially with Transformers. TL addresses these issues by providing more robust PD diagnosis and better generalizability across languages. DAFE aims to improve the explainability and interpretability of results by examining the specific effects of deep features on both other DL approaches and more traditional machine learning (ML) methods. However, it often underperforms compared to E2E and TL approaches.

**Keywords:** Parkinson's disease; speech classification; deep learning; speech processing; speech diagnosis; Artificial Intelligence; end-to-end; transfer learning; deep acoustic feature extraction

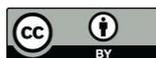

## 1. Introduction

Parkinson's disease (PD) is a neurodegenerative disorder that affects more than 10 million people around the world, being the second most common neurodegenerative disease after Alzheimer's [1]. People are generally diagnosed around the age of 65 years old [2]. Some of the symptoms include tremors, rigidity, and slowness of movement. These symptoms tend to appear in later stages, making early diagnosis based solely on them difficult [3]. PD can affect various aspects of speech production, including articulation, phonation, prosody, and voice quality [4], causing speech impairments characterized by dysphonia [5] and dysarthria [6]. Dysphonia implies a decrease in the ability to produce vocal sounds, while dysarthria refers to difficulty in producing words. These speech impairments can occur up to five years before other symptoms, suggesting that early PD diagnosis might actually be possible [7].

Impaired speech is not always easy to recognize for humans at an early stage of the disease, however, a considerable number of studies have shown the potential of automated speech-based methods to diagnose neurological disorders, including PD, from the speech signal even at the most early stages [7–9]. Such methods can be deployed for several speech tasks, of which speech classification and speech processing are the main ones.

Speech classification generally aims to distinguish between people with Parkinson's (PWP) and healthy controls (HC) [10]. Using speech-processing techniques, acoustic features (also known as acoustic biomarkers) can be extracted from speech signals to capture these changes. Classical features include fundamental frequency (F0), jitter, shimmer, Mel Frequency Cepstral Coefficients (MFCCs), and formant frequencies [1]. However, recent studies in the literature employ deep acoustic feature extraction (DAFE) from end-to-end (E2E) systems, auto-encoders, or pre-trained models [11–13]. Machine learning (ML) approaches require manual feature extraction, while deep learning (DL) methods automatically extract deep features [14]. Additionally, these features tend to be more abstract and robust [14], potentially leading to a more accurate diagnosis of PD. These features can be used not only in the final layers of a deep learning model but also in more traditional machine learning approaches, offering more interpretable results, albeit with potentially lower performance.

Until 2016, traditional methods such as Gaussian mixture models (GMMs) and hidden Markov models (HMMs) were the standard ML approach for speech processing tasks [15]. GMMs and HMMs typically use an acoustic, language, and pronunciation model. However, training each of these models separately and aligning their training data can be labor intensive. Recently, DL has emerged as the leading computational method in AI and its subfield, ML, delivering outstanding outcomes on a variety of complex cognitive tasks, and sometimes even outperforming humans [15,16]. In healthcare, DL has proven its potential not only in PD classification but in other clinical areas [17] such as cancer diagnosis [18], drug discovery [19], and precision medicine [20]. Studies that have been published in most recent years within the realm of speech-based DL approaches for PD classification mainly focus on E2E learning [21–23] and transfer learning (TL) [24–26].

Recently, there has been a growing interest in privacy-related issues in speech-based AI research [27,28]. Scientists increasingly question the reliability of speech-based AI systems, becoming more aware of concerns and risks over the trustworthiness of these systems since speech can be used as a biometric identifier to identify individuals and obtain information about their health status [27]. Datasets containing the speech of patients with PD are sensitive in nature, and safety and fairness (e.g., without bias) need to be ensured [27]; therefore, the number of such datasets that are publicly available is limited (see Table A1) as is the open-source code for the research experiments (see Table A3).

This review presents an overview of the most recent, innovative speech-based DL approaches for PD classification as of March 2024. Previously, other authors have reviewed different approaches to PD classification from various perspectives. Ngo et al. [1] summarized the literature from 2010 to 2021 related to voice and speech ML and DL approaches for PD classification and the assessment of its severity. Saravanan et al. [10] presented a review of papers until 2021 on ML and DL methods that are not only related to speech but also to physiological signals, such as electroencephalogram (EEG) and neuroimaging techniques, such as functional magnetic resonance imaging (fMRI) and single-photon emission computed tomography (SPECT). In this review, we focus on the most recent (from 2020 to March 2024) innovative DL-based approaches related to speech, which, to the best of our knowledge, have not yet been covered in previous literature reviews, guided by the overarching research question: *What is the current landscape of speech-based DL approaches for PD classification?* which has been further subdivided into the following:

**RQ1.** *Which recent speech-based DL approaches are being considered for PD classification?*

**RQ2.** *To what extent can speech-based DL approaches classify PD?*

**RQ3.** *What are the issues related to bias, explainability, and privacy of speech-based DL approaches for PD classification?*

The contributions of this systematic literature review are as follows:

1. This review covers the most recent speech-based DL approaches for PD classification, until March 2024, for which the available resources, capabilities, and potential limitations are discussed. The review explicitly focuses on speech-based data and DL approaches.
2. This review discusses issues relating to bias, explainability, and privacy associated with speech-based DL approaches for PD classification.
3. This review includes an overview of publicly available speech-based datasets and open-source resources for PD classification, up to March 2024.

The rest of the paper is structured as follows. In Section 2, the eligibility criteria, exclusion criteria, and search procedure are described. In Section 3, E2E learning approaches based on applied Deep Neural Network (DNN) architectures, such as Convolutional Neural Networks (CNNs), Long Short-Term Memory networks (LSTMs), and Transformers, are discussed. In Section 4, some of the most recent TL techniques are explained, and in Section 5, we describe approaches that use DAFE. Section 6 discusses the results according to the RQs and reflects on potential implications for researchers of the theoretical and practical findings. Section 7 outlines the conclusions of this work. Finally, in the first two appendices, we show a table listing the publicly available speech-based datasets for PD research (Table A1) and an overview of scientific works performing speech classification for PD following a DL approach (Table A2). In the last appendix, we present a table containing the open-source research code available from the scientific works included in this review (Table A3).

## 2. Methodology

The scientific works for this systematic review have been selected according to the PRISMA (Preferred Reporting Items for Systematic Reviews and Meta-Analysis) guidelines [29] and registered on the public PROSPERO database under the number 575954 (https://www.crd.york.ac.uk/prospero/display_record.php?RecordID=575954 (accessed on 13 July 2024)). In Figure 1, a flow chart of the selection process is presented. We started by identifying 148 scientific works, of which 33 were in the end selected for this systematic review. In Table A2, a summary of all works found for this literature review is presented, the public datasets employed in such works are described in Table A1, and their open-source resources available are listed in Table A3.

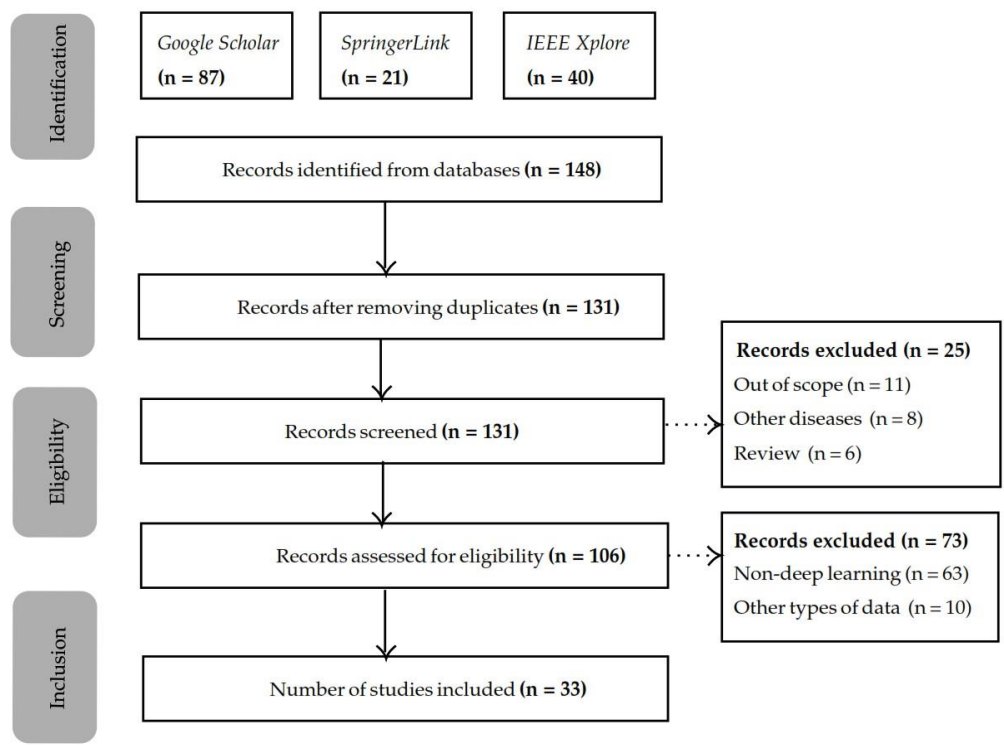

**Figure 1.** Overview of the PRISMA systematic literature review process followed.

*2.1. Search Procedure and Identification Criteria*

The search was performed in April 2024, using the following databases: Google Scholar, SpringerLink, and IEEE Xplore. The year filter was 2020 onwards. The search comprised the following search terms: "speech processing Parkinson", "speech recognition Parkinson", and "speech classification Parkinson". Although the main focus of this review is to identify innovative approaches for "speech classification", we decided to also include "speech processing" and "speech recognition" as initial search terms to make sure that we would not miss any studies that might be relevant to this review. During this step, we applied another filter to discard studies related to neuroimaging since there is a vast amount of literature only focused on the diagnosis of PD using neuroimaging techniques associated with speech impairments. The publications found were then identified by manually reading the title, abstract, and keywords, looking for terms such as deep learning, Transformers, CNNs, LSTMs, transfer learning, and deep features. The first identification search generated 148 results.

*2.2. Screening and Exclusion Criteria*

After discarding all the duplicates from the databases, there were 131 studies left. Studies that utilize traditional methods (GMMs and HMMs) or ML without DL are excluded from this review. Additionally, studies focused on neurodegenerative diseases in general, rather than specifically on PD, are also excluded. Finally, other literature reviews were discarded, leading to a total of 131 scientific words.

*2.3. Eligibility and Inclusion Criteria*

After the screening, the studies were checked on the eligibility criteria, which led to the exclusion of more studies. The main reasons for excluding these studies were the absence of deep learning (n = 63) and the presence of types of data other than speech (n = 10), while our primary interest lies in vocal features, we included works that examine both vocal and linguistic features. However, scientific studies that rely on non-speech-related features, such as gait or handwriting, fall outside the scope of this review. This led to 33 remaining studies that were taken into consideration. For each of these works, the proposed approaches and their corresponding performance were studied to gain a better understanding of the current landscape of speech-based DL approaches. From this analysis, we identified three main categories of DL approaches: E2E, TL, and DAFE. These categories are described and discussed in the following sections of this work. An overview of these categories is presented in Figure 2.

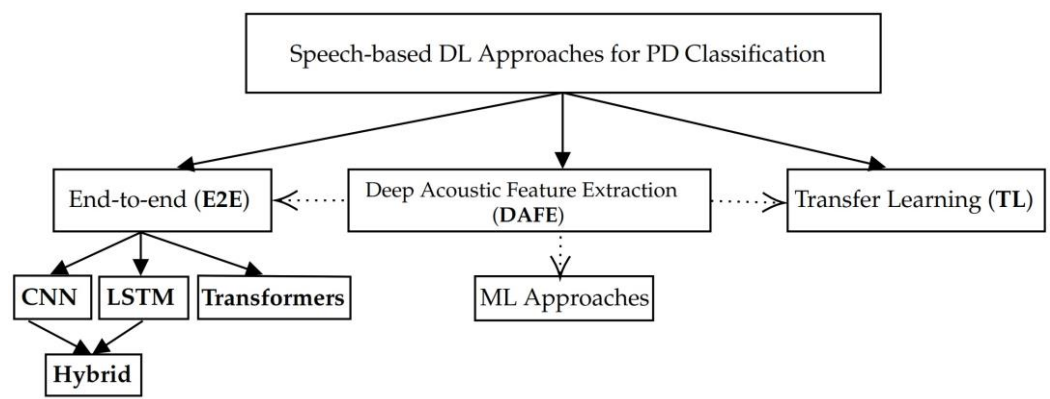

**Figure 2.** Categories of speech-based DL approaches for PD classification.

**3. E2E Approach**

In this section, we describe several studies that propose a speech-based E2E approach for PD classification. In this approach, the raw speech signal can be directly mapped to the final output [30]. This contrasts with the classical speech recognition pipeline that uses

GMMs and HMMs. In this traditional approach, handcrafted features such as MFCCs must be extracted to build an acoustic model. These features are then mapped to textual symbols using a language model before final classification can be performed [15]. We categorize the speech-based E2E approaches for DL classification into three sub-approaches: CNNs and LSTMs, hybrid CNN-LSTM architectures, and Transformers.

*3.1. CNNs and LSTMs*

CNNs are primarily designed for processing images and generally consist of convolutional, pooling, and fully connected layers [31]. Spectrograms, which represent the frequency spectrum of audio signals over time as an image, can be given as input to a CNN to identify patterns that are indicative of PD [32]. An LSTM is a type of Recurrent Neural Network (RNN) designed to capture long-range dependencies. Traditional RNNs usually use only one hidden state maintaining an internal memory, while LSTMs typically consist of an input gate, an output gate, and a forget gate [33]. This allows the LSTM to learn which information is relevant over longer sequences, which are also contained in the audio signals obtained from PD patients [34]. Khaskhoussy & Ayed [35] implement an SVM and LSTM classifier and applied these to five different datasets. The LSTM classifier consistently obtained a better F-score on all datasets (99.0%, 99.0%, 81.0%, 83.0% and 94.0%, respectively) than the SVM classifier (98.0%, 97.0%, 68.0%, 66.0% and 69.0%, respectively). The authors also report a significant difference in performance between men and women PD classification in favor of the masculine gender.

Quan et al. [21] introduce an E2E method consisting of two modules. The first module contains a series of time-distributed 2D-CNN blocks to transform the input to time series dynamic features. In the second module, a 1D-CNN block is built to learn the dependencies between these features. The authors preferred this 1D-CNN over an RNN architecture (e.g., LSTM), despite the fact that RNNs have been specifically developed for sequential input. For each single element in a given sequence, RNNs consider not only that specific element, but also all preceding elements in the sequence. However, time dynamic features are often locally correlated, resulting in local motifs that are more easily detected by 1D-CNNs than RNNs. Quan et al. [21] test their method against two databases: a Chinese database collected at the GYENNO SCIENCE Parkinson's Disease Research Center, and PC-GITA, which is a Spanish database originally presented in [36]. On both databases, the proposed method outperforms traditional ML methods such as Support Vector Machines (SVM), K-Nearest Neighbours (KNN), and Decision Tree (DT), and reaches accuracy scores of between 66.67% and 73.33%. The CNN is able to reach an accuracy of 81.6% and 75.3% for sustained vowels and short sentences in Chinese, respectively. On the Spanish database, consisting of vowels, words, and sentences, an accuracy as high as 92% is achieved.

To increase the interpretability of their E2E approach, Quan et al. [21] apply a common feature visualization method specifically developed for CNNs, Grad-class activation maps (Grad-CAM) [37], and show that their CNN tends to focus on the lower frequency regions rather than high-frequency regions of the log Mel-spectrogram. They apply the CNN again using low-frequency (ranging from 0 to 32), high-frequency (ranging from 0 to 64), and full-frequency Mel-spectrograms, as input. The best Area Under the Curve (AUC), for both Chinese and Spanish, is obtained when the lower frequency regions were used as input, indicating that lower frequencies are more influential than higher frequencies. The AUC indicates the area under the Receiver Operating Characteristic (ROC) curve of a classification model [38]. The ROC shows the relationship between the True Positive Rate and False Positive Rate of the model at several thresholds.

Similar to Quan et al. [21], Akila et al. [23] indicate that not all features are necessary to achieve high performance. The authors introduce a MASS-PCNN, which combines a multi-agent salp swarm (MASS) algorithm with a novel PD classification neural net- work (PCNN). The MASS algorithm, an updated version of SSA, selects relevant features based on various patterns. These selected features are then input into the PCNN, which includes several convolutional and pooling layers, an inception module, and a squeeze-and-excitation (SE) module. The pooling layers summarize the most important features, the SE module enhances feature importance, and the inception module captures multi-scale data. The authors argue that these additions make their CNN architecture more discriminative, achieving an accuracy of 95.1%.

In many other studies, CNNs have been employed for PD classification [3,26,30,39–44]. In [42], Zhang et al. report an accuracy of 91%. Narendra et al. [30] implement both a traditional and E2E pipeline approach, leading to an accuracy of 67.93% for the former and 68.56% for the latter. Although the E2E approach is able to outperform the more traditional approach, the performance is considerably lower than that seen in other studies. In [39], an Acoustic Deep Neural Network (ADNN), Acoustic Deep Recurrent Neural Network (ADRNN), and Acoustic Deep Convolutional Neural Network (ADCNN) are distinguished. The authors demonstrate that the CNN architecture, ADCNN, slightly outperforms the more regular DNN and RNN (99.92% versus 98.96% and 99.88%, respectively). In [40], the authors also compare a CNN and LSTM classifier using MFCCs and Gammatone Cepstral Coefficients (GTCCs) on several subsets of the PC-GITA [36] and the Parkinson's Disease Classification [45] datasets. The experiments demonstrate that the CNN significantly outperforms the LSTM classifier, and the GTCCs provide greater accuracy compared to the MFCC, reaching an accuracy of 100%, which implies none of the PWP and HC are misclassified.

In [26], the authors use both CNN and Transformer (see Section 3.3 for more details) architectures. In the case of CNNs, they use MobileNetV2, DenseNet201, DenseNet169, DenseNet121, ResNet152, ResNet50, GoogleNet, VGG19, and VGG16. The lowest accuracies are reported for the MobileNetV2 and VGG19, which led to an accuracy of 95.61%, and the highest accuracy is reported for ResNet152, for which 98.08% accuracy is reported. In this study, the authors use the Synthetic vVowels of Speakers with PD and Parkinsonism dataset [46], about which more information can be found in Table A1.

Finally, other studies on the performance of CNN-based auto-encoders have demonstrated varying levels of accuracy across different databases, privacy-aware methods, and noisy environments. Sarlas et al. [43] report an accuracy of 61.49% on the MDVR-KCL database following a Federated Learning (FL) method; whereas Janbakhshi et al. [44] observe a higher accuracy of 75.4% on the PC-GITA dataset [36] using speaker identity invariant representations and adversarial training. In Faragó et al. [41], it is pointed out that CNNs can still reach an accuracy ranging from 92% to 96% in noisy environments on a custom private small dataset of 27 participants. In Hireš et al. [3], the authors demonstrate that adding noise can even improve the accuracy with which PD can be classified on the PC-GITA dataset.

*3.2. Hybrid CNN-LSTM Architectures*

While CNNs typically outperform LSTMs in PD classification tasks, some studies indicate that hybrid CNN-LSTM architectures can also perform effectively. Er et al. [32] use variational mode decomposition to denoise speech signals and obtain the Mel-spectrograms. Next, pre-trained ResNets models (ResNet-18, ResNet-50 and ResNet-101) are used to extract deep features from the Mel-spectrograms, which are then given as input to the LSTM model. The best result is obtained when combining the ResNet-101 and LSTM, leading to an accuracy of 98.61%.

Mallela et al. [47] use four different types of speech task stimuli: subjects describing images shown to them on a computer screen in their native language (IMAG), sustained phonemes (PHON), monosyllabic targets (DIDK), and spontaneous speech (SPON). The authors apply stimuli-specific and pooled models to each of these stimuli. The stimuli-specific models are trained and tested on one specific speech task stimulus (e.g., train on IMAG and test on IMAG), whereas the pooled models are trained on all types of speech tasks stimuli and tested on each stimulus individually. For both the stimuli-specific and pooled model, the highest accuracies (96.20% and 98.27%) are achieved when using SPON, followed by IMAG, DIDK, and PHON. For PHON, the accuracy scores are considerably lower than for the other speech stimuli, ranging from 59.45% to 73.57% and 70.25% to 82.18% for the stimuli-specific and pooled model, respectively. It can be observed that the pooled model outperforms the stimuli-specific model in all cases. This suggests that the hybrid CNN-LSTM benefits from receiving various types of speech tasks as input during training. In another study performed by these authors [48], they again employ a hybrid CNN-LSTM architecture and are able to reach an accuracy of 90.98%. Initially, this model is trained to predict three classes, Amyotrophic Lateral Sclerosis (ALS), PD, and HC, and is then fine-tuned to perform binary classification between PD and HC only.

*3.3. Transformers*

Transformers were originally developed for Natural Language Processing (NLP) applications [49]. Unlike RNNs and CNNs, Transformers do not use recurrent and convolutional mechanisms. Instead, Transformers use the so-called attention mechanism which allows for parallelization. In several studies [50–52], the Transformer architecture has been adopted for PD diagnosis. Nijhawan et al. [53] implement a Vocal Tab Transformer consisting of a feature embedder, Transformer encoder, and Multilayer Perceptron (MLP) head. They name their proposed method Vocal Tab Transformer, because they obtain dysphonia measures, i.e., tabular vocal features, from the subjects' voice recordings. An XgBoost model is trained to estimate the importance of these features, of which only the most important ones are given as input to the feature embedder block. However, one of the main advantages of DNN architectures such as Transformers is that they can just take in the voice recordings as raw input [14], instead of performing a feature extraction and feature selection process, first. Although Nijhawan et al. [53] are able to reach an AUC of 0.917, other studies (e.g., [50,51]) show even better performance, without the need to implement any additional feature extraction and selection steps.

In [50], Chronowski et al. use Wav2vec2.0 [54] as a backbone model and reach an accuracy of 97.92% for the classification of PD and healthy control subjects. The authors also attempt to predict the severity of the disease based on the Hoehn and Yahr (H&Y) scale. This attempt is less successful. Unfortunately, the authors do not mention the actual performance for this task, making it difficult to determine the extent of their lack of success.

Malekroodi et al. [51] reach an accuracy of 98.5% using a Swin Transformer for the classification of PWP and HC. The margin with the CNN architecture they implemented, a VGG16, is not that large, however. With the VGG16, an accuracy of 98.1% was achieved, which is only 0.4 percentage point lower than the accuracy that was reached with the Swin Transformer. Nevertheless, it shows the potential of Transformer architectures for PD classification. Furthermore, the authors discriminate between HC, mild PD, and severe PD, based on the Unified Parkinson's Disease Rating Scale (UPDRS). They show that the Swin Transformer has strong capabilities to discriminate between HC and PD, showing a precision of 95%. However, discriminating between only two classes, mild and severe PD, appears to be more difficult, resulting in a precision of 85%. Although the recognition of different stages of PD is still not optimal, there already seems to be some improvement compared to the unsuccessful attempt to discriminate between five different classes in [50].

In Section 3, it was mentioned that [26] use several CNN and Transformer models to classify PD. For each of these models, the authors use also the TL approach, which will be discussed in more detail in Section 4. The Transformer models are the ViT-L-32 and ViT-B-16, which are both Vision Transformers (ViT), and lead to an accuracy of 96.74% and 91.57%, respectively. For most of the CNN models, such as ResNet50 (see Section 3 for more details), a higher performance was reported than for these Transformers, although the margin is small. The authors conclude from this that the latest and more advanced models are not necessarily better in terms of performance than the more conventional ones.

**4. TL Approach**

TL allows the transfer of knowledge from a source domain to a target domain [55]. In most cases, a DNN is pre-trained on a large-scale database for a certain task, which we refer to as the source domain, and then re-trained for another task usually referred to as the target domain [55]. Karaman et al. [24] use three different architectures (SqueezeNet1_1, ResNet50, and DenseNet161) that have already been trained on ImageNet [56], a database consisting of more than 14 million images. ResNet and DenseNet are dense architectures while SqueezeNet is a lightweight architecture with fewer parameters [24]. At first, the authors freeze the convolutional layers of the three architectures, and only the fully connected layers are retrained to explore the learning rate from $1.0 \times 10^{-6}$ to 1.0. Then, the convolutional layers are unfrozen and the learning rate dynamically updated. With a dynamically updated learning rate, the minimum loss can be determined more accurately. Of the three architectures, DenseNet161 obtained the highest accuracy (89.75%). Furthermore, the DenseNet-161 obtained a test time of 0.029 ± 0.028 and a training time of 0.003 ± 0.005, which is satisfactory and acceptable for clinical practice.

Hires et al. [5,57] use the German Saarbruecken Voice Database [58] and the Colombian Spanish PC-GITA dataset as their main databases [36] in their cross-lingual studies. More information about both datasets can be found in Table A1. A cross-lingual architecture is pre-trained on one certain language, also called the base language, and is then applied to another language, which is often referred to as the target language. In [5], an ensemble of CNN networks is introduced using a Multiple Fine-Tuning (MFT) approach. The ensemble consists of three base CNN networks, of which each network is pre-trained on ImageNet but fine-tuned on a different dataset. The first CNN network is fine-tuned on PC-GITA whereas the second CNN network is fine-tuned on the Vowels dataset [59], and then further fine-tuned on PC-GITA. The third CNN network is fine-tuned on the Saarbruecken Voice Database and PC-GITA datasets. The Vowels dataset is not included in Table A1 since it does not contain any information about pathological diseases [5]. The CNN architectures used in this study are the ResNet50 and Xception networks. In general, fine-tuning these two types of CNNs on multiple databases (MFT) seems to boost the performance. For Xception, using MFT is associated with a higher accuracy in three of the five vowel utterance tasks. For ResNet50, this is the case for even four of these tasks. Furthermore, the ensemble of multiple fine-tuned CNNs generally performs better than the multiple fine-tuned CNNs individually. This shows the potential of both the MFT and ensemble approach proposed by these authors. In [57], Hires et al. train and test a CNN network on a single dataset, unseen data and a mix of datasets. These datasets include the private Czech Parkinsonian dataset (CzechPD) [60], the PC-GITA dataset [36], the Italian Parkinson's voice and speech dataset [61], and the RMIT-PD dataset [62]. The authors of the experiment describe that the CNN performed well on each a single dataset, showing an accuracy of 90.82%, 90.52%, 97.81%, and 94.83% for the CzechPD, PC-GITA, Italian dataset, and RMIT-PD, respectively. However, when the CNN is tested on unseen data or a mix of the different datasets, the performance decreases significantly, leading to accuracy scores of between 43.07% and 78.85%.

Compared to Hires et al. [57], Vasquez-Correa et al. [25] are able to obtain slightly better results with their TL approach. They obtained accuracy scores of between 61.5% and 77.3% with either Czech, German, or Spanish datasets as a base language and one of the remaining two as a target language. When the Spanish dataset is used as a base language, this generally led to higher performance (77.3% and 72.6%) than when Czech or German (70.0%, 72.0%, 76.7%, 70.7%) is used as a base language. This is in line with an earlier work by these authors [63], in which they demonstrate that the accuracy increased from 69.3% to 77.3% for German and from 68.5% to 72.6% for Czech while using the Spanish dataset as the baseline.

Orozco-Arroyave et al. [64] also observe that the Spanish dataset is generally a better baseline than the Czech and German datasets for PD classification. When Spanish is the base language, accuracy scores of around 90% are reached. When Czech is the base language, accuracy scores are lower and vary within a considerably wider range (60 to 80% accuracy). Specifically, these researchers experiment with moving fractions of the target language to the training data, and excluding these fractions, then, from the test data. Only 30% has to be moved when German is the target language and Spanish is the base language to obtain an accuracy of approximately 90%. However, when Czech is the base language, the accuracy drops to 60%.

Arasteh et al. [65] employed an FL approach on specifically PC-GITA (Spanish), a German dataset with 176 subjects (88 PD and 88 HC), and a Czech dataset with 100 subjects (50 PD and 50 HC). They utilized a model consisting of four fully connected layers (1024, 256, 64, and 2) with Wav2vec2.0 embeddings. The model achieved a high robustness across different languages and populations with accuracies of 83.2%, 78.9%, and 77.8% on the Spanish, German, and Czech datasets, respectively.

Finally, there are other works in the literature which deal with the classification of several diseases using the same approach. In [25], the authors combine CNNs with TL among different diseases. A base model is trained on either Parkinson's or Huntington's disease (HD), and then the model is tested on the other disease. The CNNs are more accurate in distinguishing between HD and non-HD, than PD and non-PD. Not surprisingly, the models pre-trained on HD tend to be more accurate than models that are pre-trained on PD. This shows again that some base models transfer better than others. In other words, similar to what is observed for TL among languages [63], HD seems to transfer better than PD [25]. Another study that uses TL among diseases is [42], which has already been mentioned in Section 3.2. Although several studies take a cross-pathological approach, more studies in the existing literature are cross-lingual. These cross-lingual studies are not limited to pre-trained DNN architectures and TL applications. For studies focused on deep feature extraction, a cross-lingual approach also seems to becoming more popular. In Section 5, it is discussed in greater detail.

## 5. DAFE Approach

DAFE refers to deep features that are automatically extracted from an audio signal using a DL model. These features are either learned in the final layers of the DL model or used as input for a ML model [11]. There are a few recent studies that employ DAFE for PD classification using different DL and ML approaches. For instance, Karan et al. [66] propose a stacked auto-encoder DNN framework to classify PD and HC using speech material from the PC-GITA database. They achieve an accuracy of 87% using time–frequency deep features from spectrograms and a softmax deep classifier in the final layer, outperforming a classical ML-based SVM classifier (which achieves an accuracy of 83%).

Ferrante et al. [12] aim to understand how well a classification model trained on DAFE in one language works in a different target language. To investigate this, they use three different architectures (Wav2Vec2.0, VGGish, and SoundNet) to generate the DAFE. They alternate between English and Telugu as source and target languages, with Telugu being an under-resourced language. This could affect the accuracy of PD classification in this language. However, the results in [12] show that an accuracy higher than 90% is achieved for both English and Telugu, despite Telugu being an under-resourced language. This accuracy is observed for both the traditional features (MFCCs, F0, jitter, shimmer) and the DAFE. Another study that uses DAFE in multi-lingual and cross-lingual settings is carried out by Favaro et al. [13]. They compare the performance of interpretable feature-based models (IFM) and non-interpretable feature-based models (NIFM). They also consider pitch, loudness, and variation as interpretable features, similar to Ferrante et al. [12]. Non-interpretable features are the DNN embeddings extracted with Wav2Vec2.0, TRILLsson, and HuBERT. According to the authors, their work is the first to apply the TRILLsson and HuBERT representations to PD diagnosis and Wav2Vec2.0 in multi-lingual and cross-lingual settings [13]. They show that NIFMs outperform IFMs in mono-lingual, multi-lingual, and cross-lingual settings. In the mono-lingual setting, IFMs achieve 4% lower accuracy than NIFMs. In the multi-lingual setting, where classifiers are trained on each language separately, NIFMs outperform IFMs by 7%. This improvement over the mono-lingual setting suggests that using more languages enhances PD diagnosis. In the cross-lingual setting, where classifiers are trained on all languages except the target language, NIFMs again outperform IFMs, this time by 5.8% in terms of accuracy. One of the datasets Favaro et al. [13] use is the NeuroVoz database [67], detailed in Table A1.

Jeancolas et al. [9] also use DAFE, which they call X-vectors, instead of traditional MFCCs. They demonstrate that PD detection performs better in men due to greater variability in female MFCC distributions, which complicates MFCC-based classification in women, despite the improvements obtained with X-vectors of from 7% to 15% in terms of accuracy.

Ma et al. [68] introduce deep dual-side learning, comprising both deep feature learning and deep sample learning. For deep feature learning, the authors use an embedded stack group sparse auto-encoder (EGSAE) to acquire the deep features. Then, these deep features are fused with the original features using L1 regularization. This results in a hybrid feature set. For deep sample learning, hierarchical sample spaces are created using an iterative clustering algorithm (IMC). Classification models are applied to each of these sample spaces. A weighted fusion mechanism fuses the different models into an ensemble model, thereby combining deep feature and deep sample learning. With this method, accuracy scores of 98.4% and 99.6% are obtained on the LSVT Voice Rehabilitation Dataset [69] and the Parkinson's Disease Classification dataset [45], respectively. Whereas the Parkinson's Disease Classification dataset is publicly available (see Table A1), the raw audio files of the LSVT dataset are private.

Finally, Laganas et al. [70] aim to develop a "privacy-aware" method for classifying PD between PWP and HC using running speech signals from passively captured voice call recordings. The study involves a multi-lingual cohort of 498 subjects, comprising 392 HC and 106 PWP. A key feature of this method is that the data processing is performed locally on the smartphone, ensuring privacy by not transmitting sensitive information. The AUC for this classification method is 0.84 for the English sub-cohort, 0.93 for the Greek sub-cohort, and 0.83 for the German sub-cohort, demonstrating the method's efficacy across different languages.

## 6. Discussion

This section summarizes the main lessons learned from each research question, establishes connections with the existing literature, and elaborates on the implications for the research community.

### 6.1. Findings and Implications

Regarding *RQ1. Which speech-based DL approaches are considered for PD classification?* this systematic review has categorized and described the most recent scientific works in the literature about speech-based DL approaches for PD classification into three main categories, E2E, TL, and DAFE. In particular, CNN architectures under the E2E learning approach are the most popular speech-based approach for PD classification. Most of the E2E-based studies described in Section 3 show that CNNs can consistently outperform other architectures, such as LSTMs and hybrid CNN-LSTMs. However, in recent years, the Transformer architecture has gained in popularity and is able to compete with CNNs in terms of performance. As for our *RQ2. To what extent can speech-based DL approaches classify PD?* whereas CNNs tend to focus more on local areas in the spectrogram, Transformers show more widespread patterns [51]. Hemmerling et al. [52] describe that, intuitively, the features related to PD are not local and that it can, therefore, be expected that Transformers will outperform CNNs and increasingly dominate the field of speech-based PD classification in the following years. However, thus far, Transformers have not been able to outperform CNNs by a significant margin. Additionally, Transformers typically require extensive training data, which presents a challenge because PD speech datasets are scarce and tend to be relatively small. This limitation could hinder the growing popularity of Transformer-based approaches. Chronowski et al. [50] attributed their unsuccessful attempt to the insufficient number of subjects available for each severity stage identified by the H&Y scale, which consists of five categories. For instance, they had only two subjects for stage 1 and one subject for stage 5. Reducing the number of stages could potentially alleviate this issue as shown in Malekroodi et al. [51]. They demonstrate that having fewer classes (two) might result in a larger number of subjects per class, thereby facilitating the learning and prediction of PD severity by a Transformer model. Although not all attempts to predict PD severity have been successful, these efforts remain relevant, particularly for monitoring disease progression [71]. Datasets lacking information about PD severity of the patients prevent researchers from distinguishing between different stages of PD and more advanced PD research. Therefore, it is advisable to enrich datasets with such information.

As for the second DL approach, TL (explained in Section 4) has been shown to mitigate the issue of data availability. Therefore, there has been a recent increase in the number of studies utilizing TL for PD diagnosis. Most of the studies identified in this review are cross-lingual in nature, although some languages seem to transfer better than others. Several studies have shown that Spanish usually serves as a better base language for a dataset than, for instance, German or Czech (see [25,63,64]). One possible explanation for this is that PD is diagnosed most accurately in Spanish in the first place. In other words, without applying any form of TL, PD is diagnosed more accurately in Spanish compared to Czech or German [63]. Another reason is that the datasets for different languages were not elaborated under the same conditions, such as the severity stage of the disease. In more advanced stages, PD diagnosis might be easier to recognize than in earlier stages. In [64], German and Spanish patients were in more advanced stages than Czech patients. This could explain why we are better able to diagnose Spanish patients in the first place, and subsequently, why Spanish transfers better than other languages. Nonetheless, PD prediction is more accurate for Spanish than German patients while both are in more advanced stages of the disease. Therefore, it might not be entirely unreasonable to think that certain languages might transfer better than others.

While some languages transfer better than others, the growing number of cross-lingual studies is significant because a cross-lingual approach can lead to more PD classification and enhance generalizability to other languages. Ideally, a system developed for one language should be usable by speakers of other languages as well. Developing separate systems for each language is impractical, so creating a single system that is widely applicable would be highly beneficial. In [25,57], the Spanish dataset was found to be the most effective base language for development, suggesting that systems should be developed in Spanish and then adapted to other languages. The best performance reported was 77.3% in [25], 78.85% in [57], and around 90% in [64]. In [5], Hires et al. demonstrated that fine-tuning on multiple languages instead of just one improves performance. This suggests that fine-tuning models on several languages could enhance both robustness and applicability. Future research should explore which languages besides Spanish could serve as effective base languages. Current studies focus on Czech, English, German, or Italian (and Spanish), but more attention should be given to other linguistic and cultural contexts, as PD is a global disease affecting diverse populations.

However, cross-lingual studies use different datasets from different languages. These datasets may contain speech from different kinds of speech tasks, e.g., diadochokinetic (DDK) tasks, reading full sentences, and free speech. Therefore, the performance increase we often observe for cross-lingual studies might not only be attributed to the use of different languages but also to the use of different types of speech tasks. Nevertheless, using TL among different languages still proves to be useful since it can enhance the applicability of such systems in the real world. The same principle applies to other neurodegenerative diseases beyond PD: having a single diagnostic system that can be used for multiple diseases greatly enhances its applicability. Consequently, it is expected that future scientific research will increasingly be cross-lingual and cross-pathological.

In Section 5, it was shown that not all recent work focuses solely on E2E and TL approaches. DAFE (the third category of approaches) can be used not only for DL but also as input for more traditional ML methods to classify between PWP and HC and obtain a better interpretation and explainability of the results. For this approach, cross-lingual studies are becoming more common, which is similar to what we observed for the methodologies discussed in Sections 3 and 4. This underlines the potential of cross-lingual studies in terms of interpretability, explainability, and generalization. Although studies using DAFE can achieve a high performance of over 90% [12], and even obtain better results than traditional approaches like MFCCs [13], DAFE still underperforms compared to E2E and TL approaches. Therefore, it is unlikely that DAFE will dominate the field in the near future in terms of performance. However, while DAFE approaches may not be able to compete with the other two approaches, it can help to describe the explainability and interpretability of the results, an issue that should receive more attention in the future.

Finally, we cannot draw a definitive conclusion about which type of speech task is most suitable for diagnosing PD. Some of the studies reviewed in this work (e.g., [26]) achieved good performance with sustained vowels, while others performed well using sentences, such as [21]. In 2013, Sakar et al. [72] suggested that sustained vowels contain more PD-discriminative information than words and sentences. However, the best accuracy using sustained vowels was 85%, while more recent studies (e.g., [51]) have shown better performance with both sustained vowels and other types of speech. Since Sakar et al.'s study [72] was conducted over a decade ago, more traditional ML methods were used, rather than the DL methods that are of greater interest today. Kim et al. [73] argue that using a variety of speech tasks is likely to be most beneficial, particularly for assessing the severity of PD. According to these authors, while using sustained vowels minimizes acoustic variability from other sources, such a limited speech task does not provide much insight into a patient's intonation, rhythm, and articulation.

*6.2. Bias Issues*

Regarding the first issue of our last *RQ3. What are the issues related to bias, explainability, and privacy of speech-based DL approaches for PD classification?* in some studies within this systematic review such as [9,35], PD tends to be diagnosed more accurately in men than in women, which raises concerns about the trustworthiness of the research method [27]. For instance, in [35], higher F-scores are obtained for men compared to women. However, the private datasets employed in that experiment included a higher number of male PWP participants. This suggests that the higher accuracy in diagnosing PD in men may be due to the larger amount of male data available, indicating a demographic bias. Since PD is more prevalent in men than in women [74], this could explain why more men are often included in PD studies.

A gender bias is also evident in the work of Jeancolas et al. [9]. Using a DAFE approach, the authors demonstrate a noticeable difference in classification performance between men and women with PD, with men being classified more accurately than women. However, women showed a more significant intra-group performance improvement compared to men. The authors highlight that gender differences persist due to factors such as less pronounced brain atrophy and different speech neural circuits in women. Another explanation could be that women's voices typically exhibit greater variability, which can be reduced by Linear Discriminant Analysis (LDA) and Probabilistic Linear Discriminant Analysis (PLDA). The authors employed LDA and PLDA in their X-vector implementation, which may account for the greater improvement in performance for women. Nevertheless, the overall performance remained higher for men. This may be due to the fact that men and women exhibit different symptoms of PD [75]. Men tend to suffer more from rigidity and rapid eye movement behavior disorder, while dyskinesias and depression are more common in women. Additionally, men often show less verbal fluency and greater vocal monotony as a result of PD [76], making it easier to recognize PD in men using speech-based deep learning approaches. This could explain why performance is generally higher for men than for women. Future research should aim to close the performance gap between men and women so that these systems can work reliably for both genders in practice.

*6.3. Explainability Issues*

Another issue of the approaches discussed in this review is the explainability and interpretability of the results. Explainable Artificial Intelligence (XAI) is a current trending topic focused on making AI approaches more explainable [8]. Unfortunately, the explainability of all AI approaches discussed in this review tends to be a challenge. Regarding E2E learning, we can simply input the probabilistic representation of the spectrogram of the raw audio signals and receive a classification decision as a black-box output. Quan et al. [21] use Grad-CAM for feature visualization, demonstrating that low-frequency regions have a greater influence than high-frequency regions in identifying PD, as described in Section 3, while various visualization methods exist for CNNs, appropriate visualization techniques for Transformers remain limited [77]. CNNs capture spatial and local patterns, making it generally easier to visualize localized regions using methods like Grad-CAM [37]. In contrast, Transformers rely on self-attention, which is not localized, making it more challenging to create coherent and interpretable visualizations [78].

DAFE approaches may seem more explainable than E2E methods at first glance. However, in practice, we extract acoustic features from one DNN and input them into another DL layer or ML model. We still do not fully understand how these features are extracted or how they contribute to the decision of whether a person has PD. This lack of transparency means these approaches are not truly 'explainable.' Favaro et al. [13] utilize both IFMs and NIFMs, with the latter being DNN embeddings extracted using Wav2Vec2.0, TRILLsson, and HuBERT. The authors demonstrated that NIFMs outperform IFMs and traditional features such as MFCCs. Consequently, researchers will likely continue to prefer non-interpretable features over interpretable ones due to their superior performance. This underscores the importance of future work aimed at developing approaches that enhance explainability.

Enhanced explainability is essential for achieving acceptance and credibility in clinical settings [79]. As noted earlier, visualization methods are commonly used to address explainability challenges in E2E approaches, particularly in CNNs. Techniques such as deconvolutional neural networks and Deep Taylor decomposition can generate saliency maps that highlight the most relevant regions of an input image [80]. Layer-wise relevance propagation, which shows the positive or negative contribution of each pixel at every layer, may provide valuable insights for researchers, but the former two methods are likely more interpretable for clinicians. However, visualization methods are not the only solution for improving explainability. In [81], the authors introduce Listenable Maps for Audio Classifiers (L-MAC), which identify relevant audio segments. This approach could be particularly useful for PD diagnosis, as it allows researchers and clinicians to hear which parts of a sentence are most indicative of PD. Audio explanations may even be more intuitive than visual ones obtained with Grad-CAM and similar techniques.

Finally, in DAFE approaches, it is important to identify which extracted features contribute most to determining whether a person has PD. Techniques like SHapley Additive exPlanations (SHAP) can assign an importance value to each feature [82], while this insight is valuable, these features are often extracted from a DNN and are less interpretable compared to traditional features like jitter, shimmer, and MFCCs. Understanding the importance of deep features may be interesting for researchers seeking to understand a model's inner workings, but clinicians are likely more interested in the speech-related characteristics that indicate PD. Therefore, feature-importance techniques like SHAP might be less useful for speech-based PD diagnosis. Future work should focus on evaluating these systems in clinical settings, including model deployment, assessing user acceptance, and conducting a thorough cost–benefit analysis.

*6.4. Privacy Issues*

Finally, the recent DL approaches for PD classification can also lead to privacy, ethical, legal, and safety issues. This implies a scarcity of publicly available speech-based

PD datasets due to the sensitivity of medical data, challenges in standardizing and collecting high-quality speech recordings, and variability in symptoms among patients [27]. Arasteh et al. [65] explain that such approaches can improve diagnostic accuracy, but strict patient data privacy regulations often prevent institutions from sharing their data. To address this, the authors developed an FL method. FL has the advantage of not requiring institutions to share any of their data, which is crucial for protecting patient privacy. In their work, Arasteh et al. [65] focused on German, Spanish, and Czech speech data, all from separate institutions, none of which had access to the other's training data or network parameters. The institutions only had access to the aggregated network, without knowing the contributions of each institution. FL clearly performed better than using local models alone. However, the performance of FL was similar to that of centrally combining data from all the institutions. Sarlas et al. [43] also employ FL to promote more privacy-preserving methods, showing promise for scalable, privacy-preserving healthcare applications. However, the results of this study are not completely satisfactory (the highest accuracy obtained is 61.49%) and they point out the necessity of exploring more FL strategies and incorporating additional data modalities.

Janbakhshi et al. [44] use an auto-encoder and two auxiliary modules to retrieve speaker identity-invariant representations, thereby preserving privacy. The first module is an adversarial speaker identification module, and the second is a PD classifier. To obtain speaker-invariant representations, the speaker identification module is trained together with the auto-encoder in an adversarial manner. The auto-encoder is trained with both the speaker identification module and the PD classifier to classify PD based on speaker-invariant representations. The highest accuracy that was achieved with this method was 75.4%.

While the privacy-aware method developed by Laganas et al. [70] shows promising results, with AUC values of 0.84, 0.93, and 0.83 for the English, Greek, and German sub-cohorts, respectively, there remains room for improvement in classification accuracy. These results indicate that local feature extraction on smartphones can be a reliable approach for distinguishing PWP from HC, balancing both privacy and performance. However, most current studies on privacy-preserving methodologies for PD diagnosis lag behind other methods that do not prioritize privacy to the same extent. Future research should explore how DL techniques can be enhanced to be more privacy-preserving while achieving state-of-the-art performance, bridging the gap between privacy concerns and diagnostic accuracy.

*6.5. Limitations*

One of the limitations of this systematic literature review is that we have restricted the search to the following databases: Google Scholar, SpringerLink and IEEE Xplore. Similarly, this review is focused mainly on DL studies that classify PD based on speech, but we might have overlooked some articles. We tried to tackle this issue as much as possible by including "speech processing", "speech recognition" and "speech classification" as search terms. Furthermore, in some cases we decided to include studies that use both vocal and linguistic features, but terms such as "linguistic" or "language" were not used as search terms. Hence, there might be more studies using both vocal and linguistic features that could have been included here. Finally, for this review, we were interested in the most recent (from 2020 to March 2024) speech-based DL approaches for PD classification, implying that some studies that have successfully introduced some techniques for PD diagnosis, but are less recent, are not part of this review.

## 7. Conclusions

The goal of this work was to provide an overview of innovative speech-based DL approaches for PD classification. Given the rapid advancements in this field, we prioritized the most recent and innovative approaches from 33 research works published between 2020 and March 2024. This systematic literature review makes a significant contribution to the field by addressing a gap in which no comprehensive systematic review has previously been conducted. In particular, it provides a detailed overview of innovative speech-based DL approaches for PD classification, divided into three distinct DL categories. This review also discusses their available open-source resources, capabilities, and potential limitations, and highlights issues related to bias, explainability, and privacy. It also presents a list of publicly available speech-based datasets for PD. Most of the reviewed studies used private speech-based PD datasets and followed an E2E learning approach, particularly employing CNN architectures, which generally achieve the highest PD classification accuracy. However, the Transformer architecture is becoming competitive in performance; however, it requires extensive training data. Regardless, studies using TL, particularly cross-lingual, showed enhanced diagnostic system robustness and applicability across different languages, although some languages transfer better than others. While approaches utilizing DAFE achieved better explainability and reasonable performance, they often underperformed compared to E2E and TL approaches. Finally, we have analyzed bias, explainability, and privacy risks of the three speech-based DL approaches. We recommend that future studies focus on speech-based deep learning methods that minimize bias and enhance privacy and XAI, while still achieving state-of-the-art performance. Furthermore, increasing transparency by making open-source resources available will be crucial for the field.

**Author Contributions:** Conceptualization, L.v.G. and C.T.-G.; methodology, L.v.G.; formal analysis, L.v.G. and C.T.-G.; investigation, L.v.G. and C.T.-G.; resources, L.v.G.; writing—original draft preparation, L.v.G. and C.T.-G.; writing—review and editing, L.v.G. and C.T.-G.; visualization, L.v.G. and C.T.-G.; supervision, C.T.-G.; project administration, C.T.-G.; funding acquisition, C.T.-G. All authors have read and agreed to the published version of the manuscript.

**Funding:** This research was funded by the NWO research programme NGF AiNed Fellowship Grants under the project Responsible AI for Voice Diagnostics (RAIVD)—grant number NGF.1607.22.013.

**Data Availability Statement:** Not applicable.

**Acknowledgments:** We would like to thank the RAIVD team, who helped with insightful ideas and suggestions for the final version of this work.

**Conflicts of Interest:** The authors declare no conflicts of interest.

## Abbreviations

The following abbreviations are used in this manuscript:

| | |
|---|---|
| 1D | One-dimensional |
| 2D | Two-dimensional |
| ADCNN | Acoustic Deep Convolutional Neural Network |
| ADNN | Acoustic Deep Neural Network |
| ADRNN | Acoustic Deep Recurrent Neural Network |
| AI | Artificial Intelligence |
| ALS | Amyotrophic Lateral Sclerosis |
| AUC | Area Under the Curve |
| CNN | Convolutional Neural Network |
| DAFE | Deep Acoustic Feature Extraction |
| DDK | Diadochokinetic |
| DIDK | monosyllabic targets |
| DL | Deep learning |
| DNN | Deep Neural Network |
| DT | Decision Tree |
| E2E | End-to-end |
| EEG | Electroencephalogram |
| EGSAE | Embedded Stack Group Sparse Auto-encoder |
| F | Female |
| F0 | Fundamental frequency |
| FL | Federated Learning |
| fMRI | Functional Magnetic Resonance Imaging |
| GMM | Gaussian Mixture Models |
| Grad-CAM | Grad-class activation maps |
| GTCCs | Gammatone Cepstral Coefficients |
| HC | Healthy controls |
| HMM | Hidden Markov model |
| H&Y | Hoehn and Yahr scale |
| IFM | Interpretable feature-based models |
| IMAG | Subjects describing images in their native language |
| IMC | Iterative Clustering Algorithm |
| KNN | K-Nearest Neighbours |
| LDA | Linear Discriminant Analysis |
| L-MAC | Listenable Maps for Audio Classifiers |
| LSTM | Long Short-Term Memory networks |
| LSVT | Lee Silverman Voice Treatment |
| M | Male |
| MASS-PCNN | Multiple-Agent Salp Swarm Parkinson Classification Neural Network |
| MFCCs | Mel-frequency cepstral coefficients |
| MFT | Multiple Fine-Tuning |
| ML | Machine learning |
| NA | Not available |
| NIFM | Non-interpretable feature-based models |
| NLP | Natural Language Processing |
| PCA | Principal Component Analysis |
| PD | Parkinson's Disease |
| PHON | Sustained phonemes |
| PLDA | Probabilistic Linear Discriminant Analysis |
| PRISMA | Preferred Reporting Items for Systematic Reviews and Meta-Analysis |
| PSP | Progressive supranuclear palsy |
| PWP | People with Parkinson |
| RNN | Recurrent Neural Network |
| ROC | Receiver Operating Characteristic |
| RQ | Research question |
| SE module | Squeeze and Excitation module |
| SHAP | SHapley Additive exPlanations |
| SPECT | Single-Photon Emission Computed Tomography |
| SPON | Spontaneous speech |
| SSA | Singular Spectrum Analysis |
| SVM | Support Vector Machine |
| TL | Transfer Learning |
| UPDRS | Unified Parkinson's Disease Rating Scale |
| ViT | Vision Transformer |
| XAI | Explainable Artificial Intelligence |

## Appendix A

**Table A1.** Publicly available speech-based datasets for PD research. # and ↑ stand for number of (speakers) and ascending order of publication, respectively.

| Dataset (Year) ↑ | Source & References | # Speakers | Transcripts | Disease Severity | Language | Overall Speech Time | Data Quality | Speech Task(s) |
|---|---|---|---|---|---|---|---|---|
| *Saarbruecken Voice Database* (2006) | [58,83] & [5] | Total: 1002<br><br>912 PWP (F: 548, M: 454). Age: 6–94.<br><br>851 HC (F: 428, M: 423). Age: 9–84. | NA | Yes | Native German | ±300 min. | Microphone recordings at 50 kHz with 16-bit resolution | 1. Vowels (i, a, u) at normal, high, and low pitch.<br>2. Vowels (i, a, u) with rising, falling pitch.<br>3. "Sentence Guten Morgen, wie geht es Ihnen?" |
| *PC-GITA* (2014) | [36] * & [5,21,22,25,30, 32,40,44,52,57,65,66] | Total: 100<br><br>50 PWP (F: 25, M: 25). Age: 33–77.<br><br>50 HC (F: 25, M: 25). Age: 44–75. | NA | 5 to 92 on UPDRS and severity less than 4 on H&Y scale (except 2 patients) | Native Colombian Spanish | ±900 min. | Microphone recordings at 44.1 kHz with 16-bit resolution | 1. Phonation of sustained vowels.<br>2. Phonation of vowels with changing tone from low to high for each vowel.<br>3. Rapid repetition of words and phonemes (DDK).<br>4. Repetition of sets of words.<br>5. Repetition of sentences.<br>6. Reading of dialog.<br>7. Reading of sentences with emphasis on particular words. |
| *Parkinson Speech Dataset with Multiple Types of Sound Recordings* (2014) | [72,84] & NA | Total: 68 (40-training and 28-testing)<br><br>20 PWP (F: 6, M: 14). Age: 43–79.<br><br>20 HC (F: 10, M: 10). Age: 45–83. | NA | Yes | Native Turkish | NA | Microphone recordings at 44.1 kHz with 32-bits resolution | Training:<br>1. Sustained vowels (/a/, /o/, /u/).<br>2. Numbers (1 to 10).<br>3. Words.<br>4. Short sentences.<br><br>Test:<br>1. Sustained vowels (a, o). |
| *Italian Parkinson's Voice and Speech* (2017) | [61,85] & [51,57] | Total: 65<br><br>28 PWP (F: 9, M: 19). Age: 40–80.<br><br>37 HC (F: 14, M: 23). Age: 19–29 (15 speakers) and 60–77 (22 speakers). | Prompts available | Severity less than 4 on H&Y scale (except 2 patients) | Native Italian | ±116 min. | Microphone recordings at 16 and 44.1 kHz with 32-bit resolution | 1. Reading of phonetically balanced text.<br>2. Syllables /pa/ and /ta/.<br>3. Phonation of the vowels /a/, /e/, /i/, /o/, and /u/.<br>4. Reading of phonetically balanced words.<br>5. Reading of phonetically balanced sentences. |

**Table A1.** *Cont.*

| Dataset (Year) ↑ | Source & References | #Speakers | Transcripts | Disease Severity | Language | Overall Speech Time | Data Quality | Speech Task(s) |
|---|---|---|---|---|---|---|---|---|
| *Parkinson's Disease Classification* (2018) | [45,86] & [40,68] | Total: 252<br><br>188 PWP (F: 81, M: 107). Age: 33–87.<br><br>64 HC (F: 41, M: 23). Age: 41–82. | NA | NA | Turkish | NA | Microphone recordings at 44.1 kHz with 32-bit resolution | Sustained vowel /a/. |
| *Synthetic Vowels of Speakers with Parkinson's Disease and Parkinsonism* (2019) | [46,87] & [26] | Total: 83<br><br>22 PWP (F: 12, M: 10).<br><br>21 MSA (F: 12, M: 9).<br><br>18 PSP (F: 6, M: 12).<br><br>22 HC (F: 11, M: 11). | Prompts available | Yes | Native Czech | ±385 min. | Microphone recordings at 48 kHz with 32-bit resolution | Sustained vowels /a/ and /i/. |
| *Screening Parkinson's Diseases Using Sustained Phonemes* (2020) | [88,89] & NA | Total: 81 (Gender information: NA). | NA | NA | Native Turkish | ±189 min. | Microphone recordings at 48 kHz with 32-bit resolution | Sustained phonemes /a/, /o/, and /m/. |
| *NeuroVoz* (2024) | [67,90] & [13] | Total: 108<br><br>53 PWP (F: 20, M: 33) -Avg. age: 71.13.<br><br>55 HC (F: 26, M: 28, NA: 1) -Avg. age: 64.04. | Both manual and automatic transcriptions | 40 to 60 on UPDRS scale and 2 to 3 on H&Y scale | Native Castillian Spanish | ±106 min. | Microphone recordings at 44.1 kHz with 32-bit resolution | 1. 3 s of sustained vowels.<br>2. 10 s of DDK.<br>3. 16 listen-and-repeat utterances.<br>4. 30 s of free monologues. |

*\* Before downloading the dataset, you need to sign a user agreement. Please contact the first author via email to proceed.*

MDVR-KCL - https://zenodo.org/records/2867216
2019 - Jaeger H., Trivedi D. and Stadtschnitzer M. - English - 21 HC and 16 PWP - H&Y and UPDRS scales - Automatic transcriptions. Two speech tasks: one paragraph reading and spontaneous dialog. Each recording: 150 seconds.

## Appendix B

**Table A2.** Overview of studies performing speech classification for PD with DL. * stands for not publicly available (private). ↑ means items are ordered by year of publication (ascending).

| Cite | Year ↑ | Dataset | Approach | Model (s) | Performance |
|---|---|---|---|---|---|
| [48] | 2020 | Data collected at NIMHANS * | E2E, TL | 1D-CNN and BLSTM | Accuracy of 90.98% |
| [66] | 2020 | PC-GITA [36] | DAFE | Stacked auto-encoder and softmax classifier | Accuracy of 87% (time-frequency deep features from spectograms) |
| [47] | 2020 | Data collected from 60 ALS, 60 PD, and 60 HC subjects at NIMHANS * | E2E | CNN-LSTM | Highest accuracy of 88.5% for spontaneous speech |
| [42] | 2020 | Chinese dataset containing 31 PWP and 30 HC * | E2E | CNN | Best accuracy of 91% |
| [32] | 2021 | PC-GITA [36] | E2E | ResNet-101-LSTM | Highest accuracy of 98.61% |
| [44] | 2021 | PC-GITA [36] | E2E | CNN-based auto-encoder using speaker identity invariant representations and adversarial training | Best accuracy of 75.4% |
| [9] | 2021 | 221 French subjects (121 PD and 100 HC) recruited at Pité-Salpêtrière Hospital * | DAFE | X-vectors extracted from a DNN. To classify into PD and HC, for each subject these x-vectors are compared to the average x-vector for PD and HC, respectively. | X-vectors lead to higher performance than MFCCs, especially for women showing 7-15% point improvement |
| [24] | 2021 | mPower * | TL | SqueezeNet1_1, ResNet50, and DenseNet161 | Best accuracy of 89.75% with DenseNet161 |
| [35] | 2021 | Five different databases containing a varying number of subjects * | E2E | SVM and LSTM | LSTM consistently outperforms SVM, reaching F-scores around 98% for three of the databases |
| [70] | 2021 | Voice calls from 110 English, 185 Greek, and 140 German subjects using the iPrognosis aplication * | DAFE | Single-instance (SVM, Logistic Regression, RF) and MIL classifiers (NSK, STK, sMIL, mi-SVM) | 84%, 93% and 83% AUC for English, Greek and German subjects, respectively |
| [68] | 2021 | LSVT Voice Rehabilitation Dataset [69] * and Parkinson's Disease Classification dataset [45] | DAFE | Auto-encoder (EGSAE) combined with deep dual-side learning | Accuracies between 98.4% and 99.6% |
| [39] | 2021 | Telemonitoring (40 subjects) and multi-variate sound record dataset (40 subjects) * | E2E | ADNN, ADRNN, ARNN, and ACNN | CNN architecture outperforms regular DNN and RNN (99.92% versus 98.96% and 99.88%, respectively) |
| [30] | 2021 | PC-GITA [36] | E2E | Combination of CNN and MLP | Accuracy of 68.56% |
| [25] | 2021 | PC-GITA [36], PD-Czech, HD-Czech and Kay-elementrics dataset * | TL | CNN | Without TL, highest accuracy of 71.0% for Spanish. With TL, highest accuracy of 77.3% with Spanish as base language |
| [50] | 2022 | Data collected at University Hospital of the Jagiellonian University * | E2E | Pre-trained Wav2Vec2.0 conv layer and GRU layer | Accuracy of 97.92% |
| [5] | 2022 | PC-GITA [36], Saarbruecken Voice Database [58], and Vowels [59] * | E2E, TL | Ensemble of CNNs, using multiple-fine-tuning | 99% accuracy, 86.2% sensitivity, 93.3% specificity, and 89.6% AUC |
| [3] | 2022 | PC-GITA [36] | E2E | Xception | 92.33% accuracy on /a/ vowel and 82.12% accuracy on words |
| [21] | 2022 | Data collected at GYENNO SCIENCE Parkinson's Disease Research Center (30 PD and 15 HC) *, and PC-GITA [36] | E2E | 2D-CNN followed by 1D-CNN | 2D-CNN followed by 1D-CNN Accuracy of 81.6% and 75.3% on sustained vowels and short sentences in Chinese Accuracy of 92% on PC-GITA |
| [22] | 2022 | PC-GITA [36] (independent test set containing native Colombian Spanish speakers, 20 PD and 20 HC) * | E2E | 1D-CNN followed by LSTM | Accuracy of 77.5% and an F1-score of 75.3% for fusion of speech tasks (vowels, read text, and monologue) |
| [65] | 2023 | PC-GITA [36], 176 German (88 PD and 88 HC) and 100 Czech subjects (50 PD and 50 HC) * | TL | 4 fully connected layers (1024, 256, 64, and 2) using Wav2Vec2.0 embeddings | Accuracy of 83.2%, 78.9%, and 77.8% for Spanish, German, and Czech, respectively |
| [40] | 2023 | PC-GITA [36] and Parkinson's Disease Classification dataset [45] | E2E | CNN and LSTM | Highest accuracy (100%) reached for CNN on both datasets. |
| [14] | 2023 | 165 Colombian Spanish native speakers, from which 90 PWP * | E2E | 1D-CNN, 2D-CNN, and Wav2Vec2.0 | Best accuracy of 88% with Wav2Vec2.0 |
| [41] | 2023 | 3 datasets containing 16 PWP and 11 HC * | E2E | CNN | CNN Accuracy of 96% on speech energy, 93% on speech, and 92% on Mel spectrograms |

**Table A2.** *Cont.*

| Cite | Year ↑ | Dataset | Approach | Model (s) | Performance |
|---|---|---|---|---|---|
| [13] | 2023 | NeuroLogical Signals (NLS) *, Neurovoz [67], GITA *, GermanPD *, ItalianPVS *, and CzechPD * | DAFE | IFMs using prosodic, linguistic, and cognitive features. SVM, KNN, RF and XGBoost, and BG as classifiers. NIFMs using x- vectors extracted from TRILLsson, Wav2Vec2.0, and HuBERT. PLDA and PCA as classifiers | NIFMs outperform IFMs with 4%, 7% and 5.8% in mono-lingual, multi-lingual and cross-lingual settings, respectively |
| [12] | 2023 | MDVR-KCL [https://zenodo.org/records/2867216] and private Telugu dataset combined with Telugu split of Open SLR * | DAFE | Wav2Vec2.0, VGGish, and Soundnet for extracting deep acoustic features | 90% accuracy for both English and Telegu |
| [52] | 2023 | PC-GITA [36], Hungarian speech database, and Polish vowel database * | E2E | ResNet18, ResNet50, ResNext50, EfficientNet-B1, EfficientNet-B2, Swin Transformer, and Vision Transformer | F1-score of 78% for Vision Transformer. All other architectures reach F1-score ranging from 64 to 72% |
| [57] | 2023 | CzechPD *, PC-GITA [36], Italian Parkinson's Voice and Speech Database [85], and RMIT-PD * | E2E, TL | CNN and shallow XGBoost | Accuracies between 90.52% and 97.81% when training and testing on one language Accuracies between 43.07% and 72.04% when training and testing on different languages |
| [91] | 2023 | UCI Machine Learning Datasets Repository * | E2E | Acoustic DNN | R-squared of 86% |
| [53] | 2023 | UCI Machine Learning Datasets Repository * | E2E | Vocal Tab Transformer | AUC of 91.7% |
| [43] | 2023 | MDVR-KCL [https://zenodo.org/records/2867216] | E2E | CNN-based auto-encoder | Accuracy of 61.49% |
| [26] | 2024 | Synthetic vowels of speakers with Parkinson's disease and Parkinsonism [87] | E2E, TL | ViT-L-32, ViT-B-16, MobileNetV2, DenseNet201, DenseNet169, DenseNet121, ResNet152, ResNet50, GoogLeNet, VGG19, and VGG16. | Best accuracy of 98.30% with ResNet152 |
| [23] | 2024 | UCI Machine Learning Repository * | E2E | MASS-PCNN | 99.1% accuracy, 97.8% precision, 94.7% recall, and 99.5% F1-score |
| [51] | 2024 | Italian Parkinson's Voice and Speech Database [85] | E2E | VGG16, ResNet50, and Swin Transformer | Swin Transformer slightly outperformed VGG16 (98.5% vs. 98.1%) |

## Appendix C

**Table A3.** Open-source research code links related to speech-based DL approaches for PD classification. ↑ indicates that the items are ordered in ascending order by year of publication.

| Resources | Cite | Year ↑ |
|---|---|---|
| https://github.com/zhang946/Deep-Dual-Side-Learning-Ensemble-Model-for-Parkinson-Speech-Recognition (accessed on 20 July 2024) | [42] | 2020 |
| https://github.com/idiap/pddetection-reps-learning (accessed on 20 July 2024) | [44] | 2021 |
| https://github.com/jcvasquezc/DisVoice (accessed on 20 July 2024) | [25] | 2021 |
| Code available upon request. | [50] | 2022 |
| https://github.com/Neuro-Logical/speech/tree/main/Cross_Lingual_Evaluation (accessed on 20 July 2024) | [13] | 2023 |
| https://github.com/vincenzo-scotti/voice_analysis_parkinson (accessed on 20 July 2024) | [12] | 2023 |
| Code available upon request. | [51] | 2024 |

## References

1. Ngo, Q.C.; Motin, M.A.; Pah, N.D.; Drotár, P.; Kempster, P.; Kumar, D. Computerized analysis of speech and voice for Parkinson's disease: A systematic review. *Comput. Methods Programs Biomed.* **2022**, *226*, 107133. [CrossRef] [PubMed]
2. Toye, A.A.; Kompalli, S. Comparative Study of Speech Analysis Methods to Predict Parkinson's Disease. *arXiv* **2021**, arXiv:2111.10207.
3. Hireš, M.; Gazda, M.; Vavrek, L.; Drotár, P. Voice-Specific Augmentations for Parkinson's Disease Detection Using Deep Convolutional Neural Network. In Proceedings of the 2022 IEEE 20th Jubilee World Symposium on Applied Machine Intelligence and Informatics (SAMI), Poprad, Slovakia, 2–5 March 2022; pp. 000213–000218. [CrossRef]
4. Moro-Velazquez, L.; Gomez-Garcia, J.A.; Godino-Llorente, J.I.; Villalba, J.; Rusz, J.; Shattuck-Hufnagel, S.; Dehak, N. A forced gaussians based methodology for the differential evaluation of Parkinson's Disease by means of speech processing. *Biomed. Signal Process. Control* **2019**, *48*, 205–220. [CrossRef]

5. Hireš, M.; Gazda, M.; Drotár, P.; Pah, N.D.; Motin, M.A.; Kumar, D.K. Convolutional Neural Network Ensemble for Parkinson's Disease Detection from Voice Recordings. *Comput. Biol. Med.* **2022**, *141*, 105021. [CrossRef]
6. Muñoz-Vigueras, N.; Prados-Román, E.; Valenza, M.C.; Granados-Santiago, M.; Cabrera-Martos, I.; Rodríguez-Torres, J.; Torres-Sánchez, I. Speech and Language Therapy Treatment on Hypokinetic Dysarthria in Parkinson Disease: Systematic Review and Meta-Analysis. *Clin. Rehabil.* **2021**, *35*, 639–655. [CrossRef]
7. Moro-Velazquez, L.; Cho, J.; Watanabe, S.; Hasegawa-Johnson, M.A.; Scharenborg, O.; Kim, H.; Dehak, N. Study of the Performance of Automatic Speech Recognition Systems in Speakers with Parkinson's Disease. *Proc. Interspeech* **2019**, *2019*, 3875–3879. [CrossRef]
8. Junaid, M.; Ali, S.; Eid, F.; El-Sappagh, S.; Abuhmed, T. Explainable Machine Learning Models Based on Multimodal Time-Series Data for the Early Detection of Parkinson's Disease. *Comput. Methods Programs Biomed.* **2023**, *234*, 107495. [CrossRef]
9. Jeancolas, L.; Petrovska-Delacrétaz, D.; Mangone, G.; Benkelfat, B.E.; Corvol, J.C.; Vidailhet, M.; Benali, H. X-Vectors: New Quantitative Biomarkers for Early Parkinson's Disease Detection from Speech. *Front. Neuroinform.* **2021**, *15*, 578369. [CrossRef]
10. Saravanan, S.; Ramkumar, K.; Adalarasu, K.; Sivanandam, V.; Kumar, S.R.; Stalin, S.; Amirtharajan, R. A Systematic Review of Artificial Intelligence (AI) Based Approaches for the Diagnosis of Parkinson's Disease. *Arch. Comput. Methods Eng.* **2022**, *29*, 3639–3653. [CrossRef]
11. Khojasteh, P.; Viswanathan, R.; Aliahmad, B.; Ragnav, S.; Zham, P.; Kumar, D.K. Parkinson's Disease Diagnosis Based on Multivariate Deep Features of Speech Signal. In Proceedings of the 2018 IEEE Life Sciences Conference (LSC), Montreal, QC, Canada, 28–30 October 2018; pp. 187–190. [CrossRef]
12. Ferrante, C.; Scotti, V. Cross-Lingual Transferability of Voice Analysis Models: A Parkinson's Disease Case Study. In *Booklet of Abstracts–Spoken Language in the Medical Field: Linguistic Analysis, Technological Applications and Clinical Tools*; Politecnico di Milano University: Milan, Italy, 2023; pp. 40–42.
13. Favaro, A.; Tsai, Y.T.; Butala, A.; Thebaud, T.; Villalba, J.; Dehak, N.; Moro-Velázquez, L. Interpretable Speech Features vs. DNN Embeddings: What to Use in the Automatic Assessment of Parkinson's Disease in Multi-Lingual Scenarios. *Comput. Biol. Med.* **2023**, *166*, 107559. [CrossRef]
14. Escobar-Grisales, D.; Ríos-Urrego, C.D.; Orozco-Arroyave, J.R. Deep Learning and Artificial Intelligence Applied to Model Speech and Language in Parkinson's Disease. *Diagnostics* **2023**, *13*, 2163. [CrossRef] [PubMed]
15. Prabhavalkar, R.; Hori, T.; Sainath, T.N.; Schlüter, R.; Watanabe, S. End-to-End Speech Recognition: A Survey. *IEEE/ACM Trans. Audio Speech Lang. Process.* **2024**, *32*, 325–351. [CrossRef]
16. Taye, M.M. Understanding of Machine Learning with Deep Learning: Architectures, Workflow, Applications and Future Directions. *Computers* **2023**, *12*, 91. [CrossRef]
17. Silcox, C.; Zimlichmann, E.; Huber, K.; Rowen, N.; Saunders, R.; McClellan, M.; Kahn, C.N., III; Salzberg, C.A.; Bates, D.W. The potential for artificial intelligence to transform healthcare: Perspectives from international health leaders. *NPJ Digit. Med.* **2024**, *7*, 88. [CrossRef]
18. Rossin, G.; Zorzi, F.; Ongaro, L.; Piasentin, A.; Vedovo, F.; Liguori, G.; Zucchi, A.; Simonato, A.; Bartoletti, R.; Trombetta, C.; et al. Artificial Intelligence in Bladder Cancer Diagnosis: Current Applications and Future Perspectives. *BioMedInformatics* **2023**, *3*, 104–114. [CrossRef]
19. Jiménez-Luna, J.; Grisoni, F.; Weskamp, N.; Schneider, G. Artificial intelligence in drug discovery: Recent advances and future perspectives. *Expert Opin. Drug Discov.* **2021**, *16*, 949–959. [CrossRef] [PubMed]
20. Porumb, M.; Stranges, S.; Pescapè, A.; Pecchia, L. Precision medicine and artificial intelligence: A pilot study on deep learning for hypoglycemic events detection based on ECG. *Sci. Rep.* **2020**, *10*, 170. [CrossRef]
21. Quan, C.; Ren, K.; Luo, Z.; Chen, Z.; Ling, Y. End-to-end deep learning approach for Parkinson's disease detection from speech signals. *Biocybern. Biomed. Eng.* **2022**, *42*, 556–574. [CrossRef]
22. Rios-Urrego, C.D.; Moreno-Acevedo, S.A.; Nöth, E.; Orozco-Arroyave, J.R. End-to-end Parkinson's disease detection using a deep convolutional recurrent network. In *International Conference on Text, Speech, and Dialogue*; Springer International Publishing: Cham, Switzerland, 2022; pp. 326–338. [CrossRef]
23. Akila, B.; Nayahi, J. Parkinson Classification Neural Network with Mass Algorithm for Processing Speech Signals. *Neural Comput. Appl.* **2024**, *36*, 10165–10181. [CrossRef]
24. Karaman, O.; Çakın, H.; Alhudhaif, A.; Polat, K. Robust Automated Parkinson Disease Detection Based on Voice Signals with Transfer Learning. *Expert Syst. Appl.* **2021**, *178*, 115013. [CrossRef]
25. Vasquez-Correa, J.C.; Rios-Urrego, C.D.; Arias-Vergara, T.; Schuster, M.; Rusz, J.; Nöth, E.; Orozco-Arroyave, J.R. Transfer Learning Helps to Improve the Accuracy to Classify Patients with Different Speech Disorders in Different Languages. *Pattern Recognit. Lett.* **2021**, *150*, 272–279. [CrossRef]
26. Reddy, N.S.S.; Manoj, A.V.S.; Reddy, V.P.M.S.; Aadhithya, A.; Sowmya, V. Transfer Learning Approach for Differentiating Parkinson's Syndromes Using Voice Recordings. In *Advanced Computing*; Garg, D., Rodrigues, J.J.P.C., Gupta, S.K., Cheng, X., Sarao, P., Patel, G.S., Eds.; Springer: Cham, Switzerland, 2024; pp. 213–226. [CrossRef]
27. Feng, T.; Hebbar, R.; Mehlman, N.; Shi, X.; Kommineni, A.; Narayanan, S. A Review of Speech-centric Trustworthy Machine Learning: Privacy, Safety, and Fairness. *APSIPA Trans. Signal Inf. Process.* **2023**, *12*, e17. [CrossRef]
28. Rahman, W.; Lee, S.; Islam, M.S.; Antony, V.N.; Ratnu, H.; Ali, M.R.; Hoque, E. Detecting Parkinson Disease Using a Web-Based Speech Task: Observational Study. *J. Med Internet Res.* **2021**, *23*, e26305. [CrossRef] [PubMed]

29. Moher, D.; Liberati, A.; Tetzlaff, J.; Altman, D.G.T.G. Preferred reporting items for systematic reviews and meta-analyses: The PRISMA statement. *PLoS Med.* **2009**, *6*, e1000097. [CrossRef] [PubMed]
30. Narendra, N.P.; Schuller, B.; Alku, P. The detection of Parkinson's disease from speech using voice source information. *IEEE/ACM Trans. Audio Speech Lang. Process.* **2021**, *29*, 1925–1936. [CrossRef]
31. He, K.; Zhang, X.; Ren, S.; Sun, J. Deep Residual Learning for Image Recognition. In Proceedings of the IEEE Conference on Computer Vision and Pattern Recognition, Las Vegas, NE, USA, 26 June–1 July 2016; pp. 770–778. [CrossRef]
32. Er, M.B.; Isik, E.; Isik, I. Parkinson's Detection Based on Combined CNN and LSTM Using Enhanced Speech Signals with Variational Mode Decomposition. *Biomed. Signal Process. Control* **2021**, *70*, 103006. [CrossRef]
33. Bhati, S.; Velazquez, L.M.; Villalba, J.; Dehak, N. LSTM Siamese Network for Parkinson's Disease Detection from Speech. In Proceedings of the 2019 IEEE Global Conference on Signal and Information Processing (GlobalSIP), Ottawa, ON, Canada, 11–14 November 2019; pp. 1–5. [CrossRef]
34. Hochreiter, S.; Schmidhuber, J. Long Short-Term Memory. *Neural Comput.* **1997**, *9*, 1735–1780. [CrossRef]
35. Khaskhoussy, R.; Ayed, Y.B. Detecting Parkinson's Disease According to Gender Using Speech Signals. In Proceedings of the Knowledge Science, Engineering and Management: 14th International Conference, KSEM 2021, Tokyo, Japan, 14–16 August 2021; Proceedings, Part III; Springer International Publishing: Berlin/Heidelberg, Germany, 2021; pp. 414–425. [CrossRef]
36. Orozco-Arroyave, J.R.; Arias-Londoño, J.D.; Vargas-Bonilla, J.F.; Gonzalez-Rátiva, M.C.; Nöth, E. New Spanish speech corpus database for the analysis of people suffering from Parkinson's disease. In *LREC*; European Language Resources Association (ELRA): Paris, France, 2014; pp. 342–347.
37. Selvaraju, R.R.; Cogswell, M.; Das, A.; Vedantam, R.; Parikh, D.; Batra, D. Grad-CAM: Visual explanations from deep networks via gradient-based localization. *Int. J. Comput. Vis.* **2020**, *128*, 336–359. [CrossRef]
38. Huang, J.; Ling, C.X. Using AUC and accuracy in evaluating learning algorithms. *IEEE Trans. Knowl. Data Eng.* **2005**, *17*, 299–310. [CrossRef]
39. Nagasubramanian, G.; Sankayya, M. Multi-variate vocal data analysis for detection of Parkinson disease using deep learning. *Neural Comput. Appl.* **2021**, *33*, 4849–4864. [CrossRef]
40. Boualoulou, N.; Drissi, T.B.; Nsiri, B. CNN and LSTM for the classification of parkinson's disease based on the GTCC and MFCC. *Appl. Comput. Sci.* **2023**, *19*, 1–24. [CrossRef]
41. Faragó, P.; Ștefănigă, S.A.; Cordoș, C.G.; Mihăilă, L.I.; Hintea, S.; Peștean, A.S.; Ileșan, R.R. CNN-Based Identification of Parkinson's Disease from Continuous Speech in Noisy Environments. *Bioengineering* **2023**, *10*, 531. [CrossRef] [PubMed]
42. Zhang, T.; Zhang, Y.; Cao, Y.; Li, L.; Hao, L. Diagnosing Parkinson's disease with speech signal based on convolutional neural network. *Int. J. Comput. Appl. Technol.* **2020**, *63*, 348–353. [CrossRef]
43. Sarlas, A.; Kalafatelis, A.; Alexandridis, G.; Kourtis, M.A.; Trakadas, P. Exploring Federated Learning for Speech-Based Parkinson's Disease Detection. In Proceedings of the 18th International Conference on Availability, Reliability and Security, Benevento, Italy, 29 August–1 September 2023; pp. 1–6. [CrossRef]
44. Janbakhshi, P.; Kodrasi, I. Supervised Speech Representation Learning for Parkinson's Disease Classification. In Proceedings of the Speech Communication; 14th ITG Conference, Online, 29 September–1 October 2021; pp. 1–5.
45. Sakar, C.; Serbes, G.; Gunduz, A.; Nizam, H.; Sakar, B. Parkinson's Disease Classification. UC Irvine Machine Learning Repository. 2018. [CrossRef]
46. Hlavnička, J.; Čmejla, R.; Klempíř, J.; Růžička, E.; Rusz, J. Synthetic Vowels of Speakers with Parkinson's Disease and Parkinsonism [Dataset]. *Figshare*, 2019. [CrossRef]
47. Mallela, J.; Illa, A.; Suhas, B.N.; Udupa, S.; Belur, Y.; Atchayaram, N.; Ghosh, P.K. Voice Based Classification of Patients with Amyotrophic Lateral Sclerosis, Parkinson's Disease and Healthy Controls with CNN-LSTM Using Transfer Learning. In Proceedings of the ICASSP 2020−2020 IEEE International Conference on Acoustics, Speech and Signal Processing (ICASSP), Barcelona, Spain, 4–8 May 2020; pp. 6784–6788. [CrossRef]
48. Gope, D.; Ghosh, P.K. Raw Speech Waveform Based Classification of Patients with ALS, Parkinson's Disease and Healthy Controls Using CNN-BLSTM. *Proc. Interspeech* **2020**, *2020*, 4581–4585. [CrossRef]
49. Vaswani, A.; Shazeer, N.; Parmar, N.; Uszkoreit, J.; Jones, L.; Gomez, A.N.; Polosukhin, I. Attention Is All You Need. *Adv. Neural Inf. Process. Syst.* **2017**, *30*, 6000–6010. [CrossRef]
50. Chronowski, M.; Klaczynski, M.; Dec-Cwiek, M.; Porebska, K. Parkinson's disease diagnostics using AI and natural language knowledge transfer. *arXiv* **2022**, arXiv:2204.12559.
51. Malekroodi, H.S.; Madusanka, N.; Lee, B.I.; Yi, M. Leveraging Deep Learning for Fine-Grained Categorization of Parkinson's Disease Progression Levels Through Analysis of Vocal Acoustic Patterns. *Bioengineering* **2024**, *11*, 295. [CrossRef]
52. Hemmerling, D.; Wodzinski, M.; Orozco-Arroyave, J.R.; Sztaho, D.; Daniol, M.; Jemiolo, P.; Wojcik-Pedziwiatr, M. Vision Transformer for Parkinson's Disease Classification Using Multilingual Sustained Vowel Recordings. In Proceedings of the 2023 45th Annual International Conference of the IEEE Engineering in Medicine & Biology Society (EMBC), Sydney, Australia, 24–27 July 2023; pp. 1–4. [CrossRef]
53. Nijhawan, R.; Kumar, M.; Arya, S.; Mendirtta, N.; Kumar, S.; Towfek, S.K.; Abdelhamid, A.A. A Novel Artificial-Intelligence-Based Approach for Classification of Parkinson's Disease Using Complex and Large Vocal Features. *Biomimetics* **2023**, *8*, 351. [CrossRef]

54. Baevski, A.; Zhou, Y.; Mohamed, A.; Auli, M. wav2vec 2.0: A Framework for Self-Supervised Learning of Speech Representations. *Adv. Neural Inf. Process. Syst.* **2020**, *33*, 12449–12460.
55. Peng, X.; Xu, H.; Liu, J.; Wang, J.; He, C. Voice Disorder Classification Using Convolutional Neural Network Based on Deep Transfer Learning. *Sci. Rep.* **2023**, *13*, 7264. [CrossRef] [PubMed]
56. Deng, J.; Dong, W.; Socher, R.; Li, L.J.; Li, K.; Fei-Fei, L. Imagenet: A Large-Scale Hierarchical Image Database. In Proceedings of the 2009 IEEE Conference on Computer Vision and Pattern Recognition, Miami, FL, USA, 20–25 June 2009; pp. 248–255. [CrossRef]
57. Hireš, M.; Drotár, P.; Pah, N.D.; Ngo, Q.C.; Kumar, D.K. On the Inter-Dataset Generalization of Machine Learning Approaches to Parkinson's Disease Detection from Voice. *Int. J. Med Inform.* **2023**, *179*, 105237. [CrossRef]
58. Barry, W.J.; Putzer, M. Saarbruecken Voice Database. 2007. Available online: https://stimmdb.coli.uni-saarland.de/help_en.php4 (accessed on 13 July 2024).
59. Venegas, D. Vowels Dataset. 2018. Available online: https://www.kaggle.com/datasets/darubiano57/dataset-of-vowels (accessed on 13 July 2024).
60. Rusz, J.; Cmejla, R.; Tykalova, T.; Ruzickova, H.; Klempir, J.; Majerova, V.; Picmausova, J.; Roth, J.; Ruzicka, E. Imprecise vowel articulation as a potential early marker of Parkinson's disease: Effect of speaking task. *J. Acoust. Soc. Am.* **2013**, *134*, 2171–2181. [CrossRef] [PubMed]
61. Dimauro, G.; Di Nicola, V.; Bevilacqua, V.; Caivano, D.; Girardi, F. Assessment of Speech Intelligibility in Parkinson's Disease Using a Speech-To-Text System. *IEEE Access* **2017**, *5*, 22199–22208. [CrossRef]
62. Viswanathan, R.; Khojasteh, P.; Aliahmad, B.; Arjunan, S.P.; Ragnav, S.; Kempster, P.; Wong, K.; Nagao, J.; Kumar, D. Efficiency of voice features based on consonant for detection of Parkinson's disease. In Proceedings of the 2018 IEEE Life Sciences Conference (LSC), Montreal, QC, Canada, 28–30 October 2018; pp. 49–52. [CrossRef]
63. Vasquez-Correa, J.C.; Arias-Vergara, T.; Rios-Urrego, C.D.; Schuster, M.; Rusz, J.; Orozco-Arroyave, J.R.; Nöth, E. Convolutional Neural Networks and a Transfer Learning Strategy to Classify Parkinson's Disease from Speech in Three Different Languages. In Proceedings of the Progress in Pattern Recognition, Image Analysis, Computer Vision, and Applications: 24th Iberoamerican Congress, CIARP 2019, Havana, Cuba, 28–31 October 2019; Proceedings 24; Springer International Publishing: Berlin/Heidelberg, Germany, 2019; pp. 697–706. [CrossRef]
64. Orozco-Arroyave, J.R.; Hönig, F.; Arias-Londono, J.D.; Vargas-Bonilla, J.F.; Daqrouq, K.; Skodda, S.; Rusz, J.; Nöth, E. Automatic Detection of Parkinson's Disease in Running Speech Spoken in Three Different Languages. *J. Acoust. Soc. Am.* **2016**, *139*, 481–500. [CrossRef]
65. Arasteh, S.T.; Rios-Urrego, C.D.; Noeth, E.; Maier, A.; Yang, S.H.; Rusz, J.; Orozco-Arroyave, J.R. Federated Learning for Secure Development of AI Models for Parkinson's Disease Detection Using Speech from Different Languages. *arXiv* **2023**, arXiv:2305.11284.
66. Karan, B.; Sahu, S.S.; Mahto, K. Stacked auto-encoder based Time-frequency features of Speech signal for Parkinson disease prediction. In Proceedings of the 2020 International Conference on Artificial Intelligence and Signal Processing (AISP), Amaravati, India, 10–12 January 2020; pp. 1–4. [CrossRef]
67. Mendes-Laureano, J.; Gómez-García, J.A.; Guerrero-López, A.; Luque-Buzo, E.; Arias-Londoño, J.D.; Grandas-Pérez, F.J.; Godino-Llorente, J.I. NeuroVoz: A Castillian Spanish corpus of parkinsonian speech [Dataset]. *Zenodo* **2024**. [CrossRef]
68. Ma, J.; Zhang, Y.; Li, Y.; Zhou, L.; Qin, L.; Zeng, Y.; Lei, Y. Deep Dual-Side Learning Ensemble Model for Parkinson Speech Recognition. *Biomed. Signal Process. Control* **2021**, *69*, 102849. [CrossRef]
69. Tsanas, A. LSVT Voice Rehabilitation [Dataset]. *UCI Machine Learning Repository*, 2014. [CrossRef]
70. Laganas, C.; Iakovakis, D.; Hadjidimitriou, S.; Charisis, V.; Dias, S.B.; Bostantzopoulou, S.; Hadjileontiadis, L.J. Parkinson's Disease Detection Based on Running Speech Data from Phone Calls. *IEEE Trans. Biomed. Eng.* **2021**, *69*, 1573–1584. [CrossRef] [PubMed]
71. Bayestehtashk, A.; Asgari, M.; Shafran, I.; McNames, J. Fully automated assessment of the severity of Parkinson's disease from speech. *Comput. Speech Lang.* **2015**, *29*, 172–185. [CrossRef]
72. Sakar, B.E.; Isenkul, M.E.; Sakar, C.O.; Sertbas, A.; Gurgen, F.; Delil, S.; Apaydin, H.; Kursun, O. Collection and analysis of a Parkinson speech dataset with multiple types of sound recordings. *IEEE J. Biomed. Health Inform.* **2013**, *17*, 828–834. [CrossRef]
73. Kim, J.; Nasir, M.; Gupta, R.; Van Segbroeck, M.; Bone, D.; Black, M.P.; Skordilis, Z.I.; Yang, Z.; Georgiou, P.G.; Narayanan, S.S. Automatic estimation of parkinson's disease severity from diverse speech tasks. *Proc. Interspeech* **2020**, *2020*, 914–918.
74. Podcasy, J.L.; Epperson, C.N. Considering sex and gender in Alzheimer disease and other dementias. *Dialogues Clin. Neurosci.* **2016**, *18*, 437–446. [CrossRef] [PubMed]
75. Miller, I.N.; Cronin-Golomb, A. Gender differences in Parkinson's disease: Clinical characteristics and cognition. *Mov. Disord.* **2010**, *25*, 2695–2703. [CrossRef]
76. Gillies, G.E.; Pienaar, I.S.; Vohra, S.; Qamhawi, Z. Sex differences in Parkinson's disease. *Front. Neuroendocrinol.* **2014**, *35*, 370–384. [CrossRef] [PubMed]
77. Leem, S.; Seo, H. Attention Guided CAM: Visual Explanations of Vision Transformer Guided by Self-Attention. *Proc. AAAI Conf. Artif. Intell.* **2024**, *38*, 2956–2964. [CrossRef]
78. Abnar, S.; Zuidema, W. Quantifying attention flow in transformers. *arXiv* **2020**, arXiv:2005.00928.

79. Band, S.S.; Yarahmadi, A.; Hsu, C.C.; Biyari, M.; Sookhak, M.; Ameri, R.; Dehzangi, I.; Chronopoulos, A.T.; Liang, H.W. Application of explainable artificial intelligence in medical health: A systematic review of interpretability methods. *Inform. Med. Unlocked* **2023**, *40*, 101286. [CrossRef]
80. Haar, L.V.; Elvira, T.; Ochoa, O. An analysis of explainability methods for convolutional neural networks. *Eng. Appl. Artif. Intell.* **2023**, *117*, 105606. [CrossRef]
81. Paissan, F.; Ravanelli, M.; Subakan, C. Listenable Maps for Audio Classifiers. *arXiv* **2024**, arXiv:2403.13086.
82. Lundberg, S.M.; Lee, S.I. A unified approach to interpreting model predictions. *Adv. Neural Inf. Process. Syst.* **2017**, *30*, 4768–4777. [CrossRef]
83. Koreman, J.C. A German Database Of Patterns Of Pathological Vocal Fold Vibration. *Engineering* **1997**, *3*, 143–153.
84. Kursun, O.; Sakar, B.; Isenkul, M.; Sakar, C.; Sertbas, A.; Gurgen, F. Parkinson's Speech with Multiple Types of Sound Recordings [Dataset]. *UCI Machine Learning Repository*, 2014. [CrossRef]
85. Dimauro, G.; Girardi, F. Italian Parkinson's Voice and Speech [Dataset]. *IEEE Dataport*, 2019. [CrossRef]
86. Sakar, C.O.; Serbes, G.; Gunduz, A.; Tunc, H.C.; Nizam, H.; Sakar, B.E.; Tutuncu, M.; Aydin, T.; Isenkul, M.E.; Apaydin, H. A comparative analysis of speech signal processing algorithms for Parkinson's disease classification and the use of the tunable Q-factor wavelet transform. *Appl. Soft Comput.* **2019**, *74*, 255–263. [CrossRef]
87. Hlavnička, J.; Čmejla, R.; Klempíř, J.; Růžička, E.; Rusz, J. Acoustic Tracking of Pitch, Modal, and Subharmonic Vibrations of Vocal Folds in Parkinson's Disease and Parkinsonism. *IEEE Access* **2019**, *7*, 150339–150354. [CrossRef]
88. Pah, N.D.; Motin, M.A.; Kempster, P.; Kumar, D.K. Detecting Effect of Levodopa in Parkinson's Disease Patients Using Sustained Phonemes. *IEEE J. Transl. Eng. Health Med.* **2021**, *9*, 1–9. [CrossRef]
89. Kumar, D.; Kempster, P.; Raghav, S.; Viswanthan, R.; Zham, P.; Arjunan, S. *Screening Parkinson's Diseases Using Sustained Phonemes*; RMIT University: Melbourne VIC, Australia, 2020. [CrossRef]
90. Mendes-Laureano, J.; Gómez-García, J.A.; Guerrero-López, A.; Luque-Buzo, E.; Arias-Londoño, J.D.; Grandas-Pérez, F.J.; Godino-Llorente, J.I. NeuroVoz: A Castillian Spanish corpus of Parkinsonian Speech. *arXiv* **2024**, arXiv:2403.02371.
91. Mahmood, A.; Mehroz Khan, M.; Imran, M.; Alhajlah, O.; Dhahri, H.; Karamat, T. End-to-end deep learning method for detection of invasive Parkinson's disease. *Diagnostics* **2023**, *13*, 1088. [CrossRef] [PubMed]